\documentclass[preprint,12pt]{elsarticle}
\usepackage{amssymb}
\usepackage{graphicx}
\usepackage{amsmath}
\usepackage{amsfonts}
\usepackage{esint}
\usepackage{color}
\usepackage{xcolor}
\usepackage{float}
\usepackage{subcaption}
\usepackage{verbatim}
\usepackage{bbold}
\usepackage[colorlinks=true,citecolor=blue,linkcolor=blue,urlcolor=blue]{hyperref}
\usepackage[justification=centerlast,font={small}]{caption}

\newcommand{\BT}{\mathcal{T}}

\newcommand{\BP}{\mathcal{P}}

\begin{document}

\title{Quantum phase transitions 
%witohut exception points 
in non-Hermitian $\BP\BT$-symmetric transverse-field Ising spin chains}
\author{Grigory A. Starkov$^a$}
%\affiliation{Institut f\"ur Theoretische Physik III, Ruhr-Universit\"at %Bochum, Bochum 44801, Germany}
%\email{Grigorii.Starkov@rub.de}
\author{Mikhail V. Fistul$^a$}
%\affiliation{Institut f\"ur Theoretische Physik III, Ruhr-Universit\"at %Bochum, Bochum 44801, Germany}
\author{Ilya M. Eremin$^a$}
\affiliation{Institut fur Theoretische Physik III, Ruhr-Universitat Bochum, 44801 Bochum, Germany}

\date{\today}

\begin{abstract}
We present a theoretical study of quantum phases and quantum phase transitions occurring in non-Hermitian $\BP\BT$-symmetric superconducting qubits chains described by a transverse-field Ising spin model. A non-Hermitian part of the Hamiltonian is implemented via imaginary staggered \textit{longitudinal } magnetic field, which corresponds to a local staggered gain and loss terms. %respectively.  
By making use of a direct numerical diagonalization of the Hamiltonian for spin chains of a finite size $N$, we explore the dependencies of the energy spectrum, including the energy difference between the first excited and the ground states, the spatial correlation function of local polarization ($z$-component of local magnetization) on the adjacent spins interaction strength $J$ and the local gain (loss) parameter $\gamma$. A scaling procedure for the coherence length $\xi$ allows us to establish a complete quantum phase diagram of the system.
We obtain two quantum phases for $J<0$, namely, $\BP\BT$-symmetry broken antiferromagnetic state and $\BP\BT$-symmetry preserved paramagnetic state, and the quantum phase transition line between them is the line of exception points. For $J>0$ the $\BP\BT$-symmetry of the ground state is retained in a whole region of parameter space of $J$ and $\gamma$, and a system shows \textit{two} intriguing quantum phase transitions between ferromagnetic and paramagnetic states for a fixed parameter $\gamma > 1$. We also provide the qualitative quantum phase diagram $\gamma-J$ derived in the framework of the Bethe-Peierls approximation that is in a good accord with numerically obtained results. 
\end{abstract}
\maketitle

\textbf{Keywords: Quantum phase transition; Non-Hermitian $\BP\BT$-symmetric Hamiltonian; Ising spin chain; spatial correlation functions; spin polarization}

\section{Introduction}
Quantum phase transitions at zero temperature, which occur as a result of competing ground state phases with spontaneous change of macroscopic physical quantities  upon small variation of physical parameters, keep to fascinate scientific community for several decades  \cite{sachdev1999quantum,sondhi1997continuous}. The cuprate superconductors, which can be tuned from a Mott insulating to a $d$-wave superconducting phase by carrier doping are a paradigmatic example\cite{vojta2003quantum} and are still not fully understood\cite{Efetov2013}. Further examples are the quantum ferromagnetic-paramagnetic, the quantum Kosterlitz-Thouless transitions, quantum spin liquid phases have been identified in numerous condensed matter systems including but not limited to one- or two dimensional lattices of Josephson junctions \cite{sondhi1997continuous,haviland2000superconducting,ergul2013localizing}, granular metals and superconductors \cite{efetov1980phase,beloborodov2007granular,efetov2019order}, superconducting interacting qubits \cite{reiner2016emulating,johnson2011quantum,harris2018phase,king2018observation}, magnetic low-dimensional frustrated systems \cite{taillefumier2017competing} and photonic band-gap cavities \cite{greentree2006quantum}. 

A special role in the theoretical study of quantum phase transitions belongs to seminal \textit{integrable} low-dimensional models, like the quantum $X-Y$ model supporting the superconductor-insulator transition, one-dimensional spin chain systems with an exchange type of interaction displaying the Mott insulator-superfluid transition, and one-dimensional spin chains of $\vec{\sigma}=1/2$ in the presence of both an Ising type ($\hat \sigma^z_n \cdot \hat \sigma^z_{n+1}$) of interaction and a transverse magnetic field in which the quantum phase transition between paramagnetic-(anti)ferromagnetic states occurs \cite{pfeuty1970one}. In the latter case the exact eigenvalues and eigenstates can be analytically calculated by using the Jordan-Wigner transformation \cite{backens2019jordan}. However, in the presence of both transverse and longitudinal magnetic fields the system ceases to be integrable, and therefore, to obtain the quantum phase diagram one needs to use complex numerical \cite{de2019ground} or approximated analytical methods \cite{strecka2015brief,Bethe-1935,Peierls-1936,Weiss-1948}.

This field of quantum phase transitions has received the new twist as various photonic \cite{ruter2010observation,el2018non,szameit2011p} and solid state \cite{cartarius2012model,naghiloo2019,dogra2021quantum,wu2019observation} systems have been realized, whose dynamics is governed by a non-Hermitian parity-time $\BP\BT$-symmetric Hamiltonian. Such Hamiltonians can be implemented in systems where a non-equilibrium growth of the population of specially chosen quantum states, {\it i.e.}, states with a \textit{gain}, can be completely compensated by a \textit{loss }present in the other states. 

A general theoretical analysis of such systems was provided by the seminal works of C. Bender with co-workers \cite{bender1998real,bender1999pt,bender2007making}. In particular, they showed that the $\BP\BT$-symmetric non-Hermitian Hamiltonian can exhibit a purely real eigenvalues spectrum, identifying the $\BP\BT$-symmetry \textit{preserved} quantum phase. At the same time as the gain/loss parameter varies, there is also another regime of the Hamiltonian where the eigenvalues of the $\BP\BT$-symmetric Hamiltonian become complex conjugate ones, signalling the \textit{broken} $\BP\BT$-symmetric quantum phase, where the so-called exception point (line) determines the quantum phase transition between $\BP\BT$-symmetry preserved and broken phases. The phase transitions in various $\BP\BT$-symmetric systems have received some attention recently\cite{Bender2012,Jin2013,Jin2017,Ashida2017,Edvardsson2020,Nakanishi2022} yet this question was rarely addressed in the context of quantum phase transition in the known paradigmatic quantum systems. This is partially connected to the fact that there is still less known on how to realize and control $\BP\BT$-symmetric quantum Hamiltonian systems experimentally.

Recently, a general analysis have been used to identify different quantum regimes in exemplary non-Hermitian 
\textit{small} superconducting qubits (two-levels) systems, i.e. $\BP\BT$-symmetric single qubit \cite{naghiloo2019,dogra2021quantum,wu2019observation} and two interacting qubits \cite{tetling2022linear}. Therefore, a next natural question arise: what are the quantum phases and quantum phase transitions occurring in \textit{large} systems of $\BP\BT$-symmetric interacting superconducting qubits? This question was addressed in a few papers \cite{Song-14,Song-15,Schmidt-21} where Ising spin chains in a complex transverse magnetic field were theoretically studied. Notice here that such the non-Hermitian $\BP\BT$-symmetric  model is still integrable one. 

In this manuscript we present a theoretical study of the ground  and low lying exciting states occurring in Ising spin chains in the presence of both transverse and longitudinal magnetic fields. A non-Hermitian part of total Hamiltonian is provided by an \textit{imaginary } $\BP\BT$-symmetric staggered longitudinal magnetic field. This renders the presented model genuinely non-integrable.  Such an imaginary longitudinal magnetic field can be implemented in optical systems \cite{ruter2010observation,el2018non,szameit2011p}, trapped ions and ultracold atoms \cite{ding2021experimental,li2019observation}, Bose-Einstein condensate \cite{cartarius2012model}  with the $PT$-symmetric combination of gain and loss, and in superconducting \cite{naghiloo2019,dogra2021quantum} or  nitrogen-vacancies \cite{wu2019observation} qubits systems interacting with auxiliary qubits. 

We focus on the quantum phase diagram $\gamma-J$ at 
zero temperature in the thermodynamic limit of $N \rightarrow \infty$ ($N$ is a total number of spins in the chain). Here, $\gamma$ and $J$ are the gain (loss) and the coupling strength between adjacent spins, accordingly. In order to establish the quantum phase diagram we use direct numerical diagonalization of the non-Hermitian Hamiltonian for a finite size spin chain accompanied by the scaling approach \cite{pang2019critical,asada2002anderson,Sandvik-2010} and the analytical  Bethe-Peierls approximation \cite{strecka2015brief,Bethe-1935,Peierls-1936,Weiss-1948}.

%Our approaches: numerical procedure, scaling approach, %

%The scope of our work contains two topics. 
%In the first part of the paper, we consider the small number of spins %in the chain and study the behaviour of the eigenspectrum as the %strength of imaginary field is increased.
The paper is organized as follows.
In Section~\ref{section:model} we describe the model, write down the non-Hermitian $\BP\BT$-symmetric Hamiltonian for the transverse-field Ising spin chain and explain in detail our numerical procedure. 
%its analytic solution in the absence of dissipation. 
In Section~\ref{section:results} we numerically explore the dependence of the energy spectrum, the energy gap between the first excited and ground states, and a spatial correlation function of the local spin polarization ($z$-component of local magnetization) on the interaction strength $J$ and the gain (loss) parameter $\gamma$ for non-Hermitian Ising spin chains of a moderate size $N$ (up to $N=18$). In this Section we also extract the correlation length $\xi_N(J,\gamma)$, and, using the scaling analysis, obtain a complete phase diagram of a non-Hermitian $\BP\BT$-symmetric transverse-field Ising spin chain in the limit of $N \rightarrow \infty$ (see Section~\ref{section:phasediag}). In Section~\ref{section:BP} using the Bethe-Peierls approximation \cite{weiss1948application,du2003expanded} the qualitative phase diagram will be analytically reconstructed. We conclude with  Section~\ref{section:conclusions}.

%Using  consider the quantum phase diagram of the system. The diagram is %determined numerically by computing the end-to-end correlation %functions in spin chains of moderate size. In Section~\ref{section:mf}, %we provide the qualitative mean-field dedscription of the quantum phase %diagram. In Appendix~\ref{section:rules}, we formulate and prove the %existance of the symmetry-based selection rules. Finally, in the %Appendix~\ref{section:parity}, we discuss the parity of the states of %transverse-field Ising chain in the absence of the dissipation, which %is important for the application of the selection rules.

\section{Model and numerical procedure.\label{section:model}}
\begin{figure}[t]
\includegraphics[width=\textwidth]{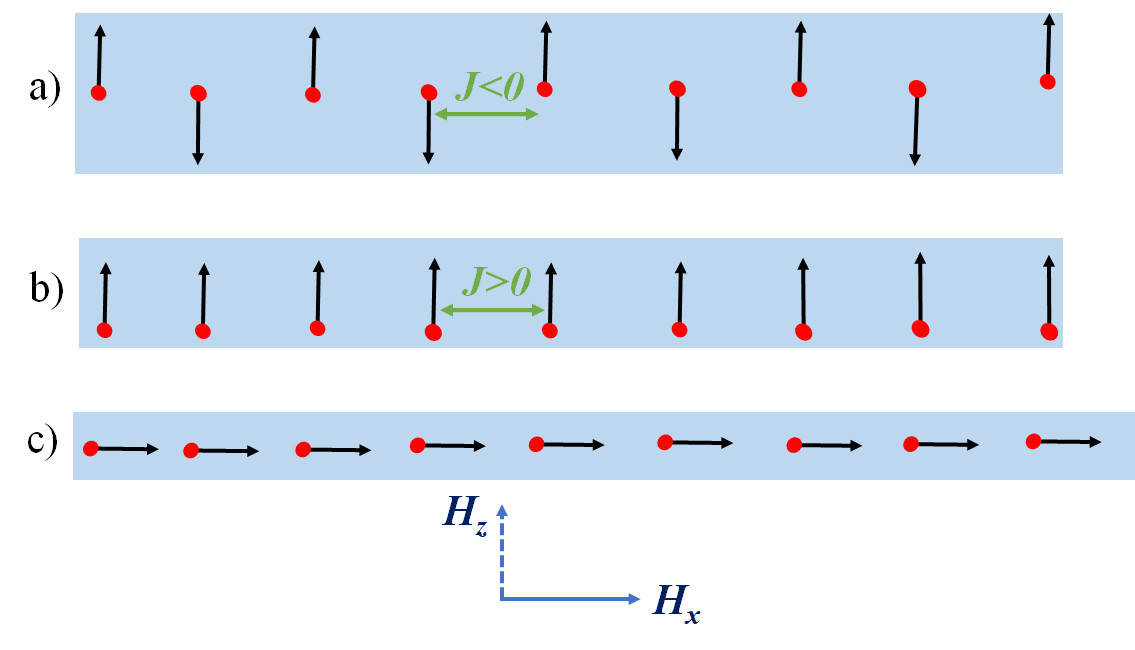}
\caption{Schematics of a one-dimensional chain of interacting Ising spins placed in a complex magnetic field. Antiferromagnetic ($J<0$) (a), ferromagnetic ($J>0$) (b) and quantum paramagnetic (c) states are shown. The transverse magnetic field, $H_x \simeq \Delta$, and an imaginary longitudinal magnetic field, $H_z \simeq i\gamma$, are indicated.}
\label{fig:model}
\end{figure}

We consider a one-dimensional chain composed of $N$ interacting spins $1/2$ placed in the transverse ($x$-direction) magnetic field $H_x$ of a strength $\Delta$. The interaction strength $J$ between adjacent spins is the Ising interaction, i.e., $-J\hat \sigma^z_n \cdot \hat \sigma^z_{n+1} $. The positive (negative) sign of $J$ determines the ferromagnetic (antiferromagnetic) couplings, accordingly.
In our model an imaginary staggered longitudinal magnetic field $H_z$ of a strength $\gamma$ provides a $\BP\BT$-symmetric non-Hermitian part of the Hamiltonian in the form, $i\gamma (-1)^n \hat \sigma^z_n $. 
%A chain size is $N$. 
The model is presented schematically in Fig. \ref{fig:model}.
At zero temperature, the physical properties of a $\BP\BT$-symmetric non-Hermitian spin chain are completely determined by the Hamiltonian written as follows
\begin{align}
    \hat{H}_{PT} &= \sum_{n=1}^N \left[\Delta\hat{\sigma}_{n}^{x}+(-1)^{n-1}\,\mathrm{i}\gamma \hat{\sigma}_{n}^{z}\right]-J\sum_{n=1}^N \hat{\sigma}_{n}^{z}\hat{\sigma}_{n+1}^{z}.
    \label{Ising-Ham}
\end{align}
To identify the quantum phases and the phase transitions as a function of the system parameters we use the following numerical procedure. Fixing the parameters $J/\Delta$ and $\gamma/\Delta$ we numerically diagonalize the Hamiltonian (\ref{Ising-Ham}) for spin chains of various sizes ($N=2 \div 18$), and obtain the eigenvalues, $\varepsilon_n$, and right (left) eigenvectors  $|R_n\rangle$ ($|L_n\rangle$). The periodic boundary conditions were used in most of numerical calculations.
Notice here, that real and complex conjugate  values of $\varepsilon_n$ determine the $\BP\BT$-symmetry preserved and broken regimes, accordingly. 

Using these eigenvalues and eigenvectors we calculate the energy difference (the energy gap) between the first excited and the ground states, $\Delta \varepsilon$, and the ground state spatial correlation function of a local spin polarization ($z$-component of a local magnetization) defined as
\begin{align}
    C(n-m) & =\langle R_0|\hat\sigma_n^z\hat\sigma_m^z|R_0\rangle.
    \label{corrfunction}
\end{align}
Here, $|R_0\rangle$ is the eigenvector of the ground state. In the $\BP\BT$-symmetry broken regime the state $|R_0\rangle$ corresponds to the eigenvalue $\varepsilon_0$ with a negative imaginary part. After that we vary the parameters $J/\Delta$ and $\gamma/\Delta$ in wide regions, i.e., $-1.5< J/\Delta < 1.5$ and $0< \gamma/\Delta < 2$, and the numerical procedure was repeated. 

\section{Results\label{section:results}}
In this Section we present our main results, i.e, the dependence of the energy spectrum, the energy gap and the spatial correlation function $C(n-m)$ on the parameters $J/\Delta$ and $\gamma/\Delta$.
\subsection{Energy spectrum}\label{section:spectrum}
\begin{figure}[p]
\includegraphics[width=\textwidth]{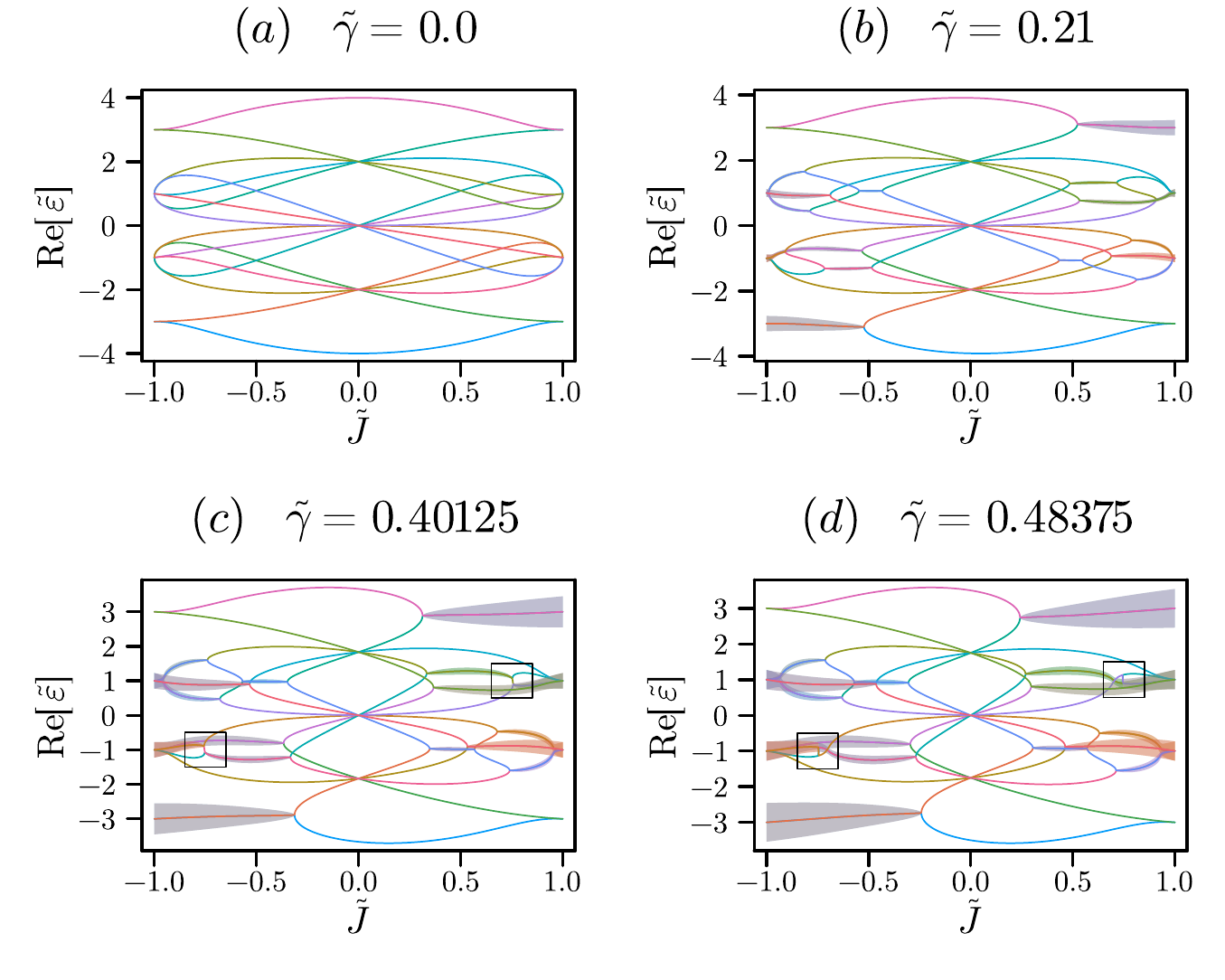}
\caption{The dependence of a real part of normalized eigenvalues $\tilde{\varepsilon}_n=\varepsilon_n/\sqrt{J^2+\Delta^2}$ on the normalized coupling strength $\tilde J=J/\sqrt{J^2+\Delta^2}$ for a $N=4$ spin chain. The parameter $\tilde\gamma=\gamma/\sqrt{J^2+\Delta^2}$ was chosen as $0; 0.21; 0.40125; 0.48375$. The shaded ribbons around the curves depict the scaled imaginary part of the normalized eigenvalues. 
In the panels (c) and (d), small open rectangles mark the positions of the third order exception points.}
\label{st4}
\end{figure}

Using the numerical procedure described in the previous Section we obtain the dependence of eigenvalues of the Hamiltonian (\ref{Ising-Ham}) on the effective coupling strength $\tilde J=J/\sqrt{J^2+\Delta^2}$ for different values of $\tilde\gamma=\gamma/\sqrt{J^2+\Delta^2}$. The typical results are presented in Fig. \ref{st4} for an exemplary spin chain with $N=4$ and open boundary conditions. The energy levels of a $\BP\BT$-symmetric Hamiltonian are either real, or form complex conjugated pairs.

From this figure one can conclude that there is an important difference in the properties of the ground state, i.e., the state with the minimal value of a real part of the eigenvalues. Indeed, for an antiferromagnetic coupling ($J<0$) the presence of gain (loss) $\gamma$ induces a  transition between the $\BP\BT$-symmetry preserved and broken regimes. Moreover, a critical coupling strength $J_{cr}$ determining a transition (the so-called  exception point) decreases with $\gamma$.   
These features are a direct consequence of the same parity symmetry of an antiferromagnetic state and a staggered imaginary longitudinal magnetic field.
In the opposite case of a ferromagnetic coupling ($J>0$) the ground state is in the $\BP\BT$-symmetry preserved regime for all values of $J$ and $\gamma$.
%These features are direct consequence of the presence of %two parity symmetries: , and odd (even) 

For the larger values of $\tilde\gamma$,
the energy spectrum develops more complicated features.
For example, the two exception points of the second order can coalesce and go through the exception point of the third order~\cite{Somnath-20,Bergholtz-21} as seen in Fig.~\ref{st4}(c,d) (indicated by small open rectangles).
Since the detailed discussion of the energy spectrum properties is a separate interesting issue we will present it elsewhere. 

%The detailed discussion of the energy spectrum properties %is a separate interesting issue but for the sake of the %presentation the details of this analysis will be %presented elsewhere.

%of The dependence of eigenvaluesAlmost all of the results in the paper %can be easily understood if one envokes the notion of the %symmetry-based selection rules, which govern what states are allowed to %form complex conjugated pairs as the strength of the imaginary field is %turned up. Stated simply, only the states with different parity can %form such pairs. Strictly speaking, the parity is defined only in the %absence of dissipation, but we show how this notion can be extended to %the $\BP\BT$-symmetric phase of the non-Hermitian Hamiltonian.

%\begin{figure*}[h]
%\includegraphics[width=\textwidth]{ep3.pdf}
%\caption{The structure of the exception point of the third %order. The plots show the real part of the normalized %eigenvalues as functions of the normalized imaginary field %strength $\tilde\gamma = \gamma/\sqrt{J^2+\Delta^2}$ for %different fixed values of normalized coupling constant %$\tilde J = J/\sqrt{J^2 + \Delta^2}$.
%The shaded ribbons depict the scaled imaginary part of the %normalized eigenvalues.
%Only the levels forming the exception point are plotted. %The annotations specify the parities of the states.}
%\label{ep3}
%\end{figure*}

\subsection{Energy gap between the first excited and ground states. \label{subsection:energygap} }
An important physical characteristics of quantum phases and quantum phase transitions is the energy gap $\Delta \varepsilon $ between the first excited and the ground states. For example, in the thermodynamic limit $T=0$ and $N \rightarrow \infty$ the quantum phase transition in the Hermitian transverse field Ising model occurs as $\Delta \varepsilon=0$ \cite{pfeuty1970one}. Moreover, such energy gap determines the resonant response of quantum systems to an applied small alternating perturbation \cite{tetling2022linear,blais2007quantum,jung2014multistability,fistul2022quantum}. 

\begin{figure}[t]
\includegraphics[width=\textwidth]{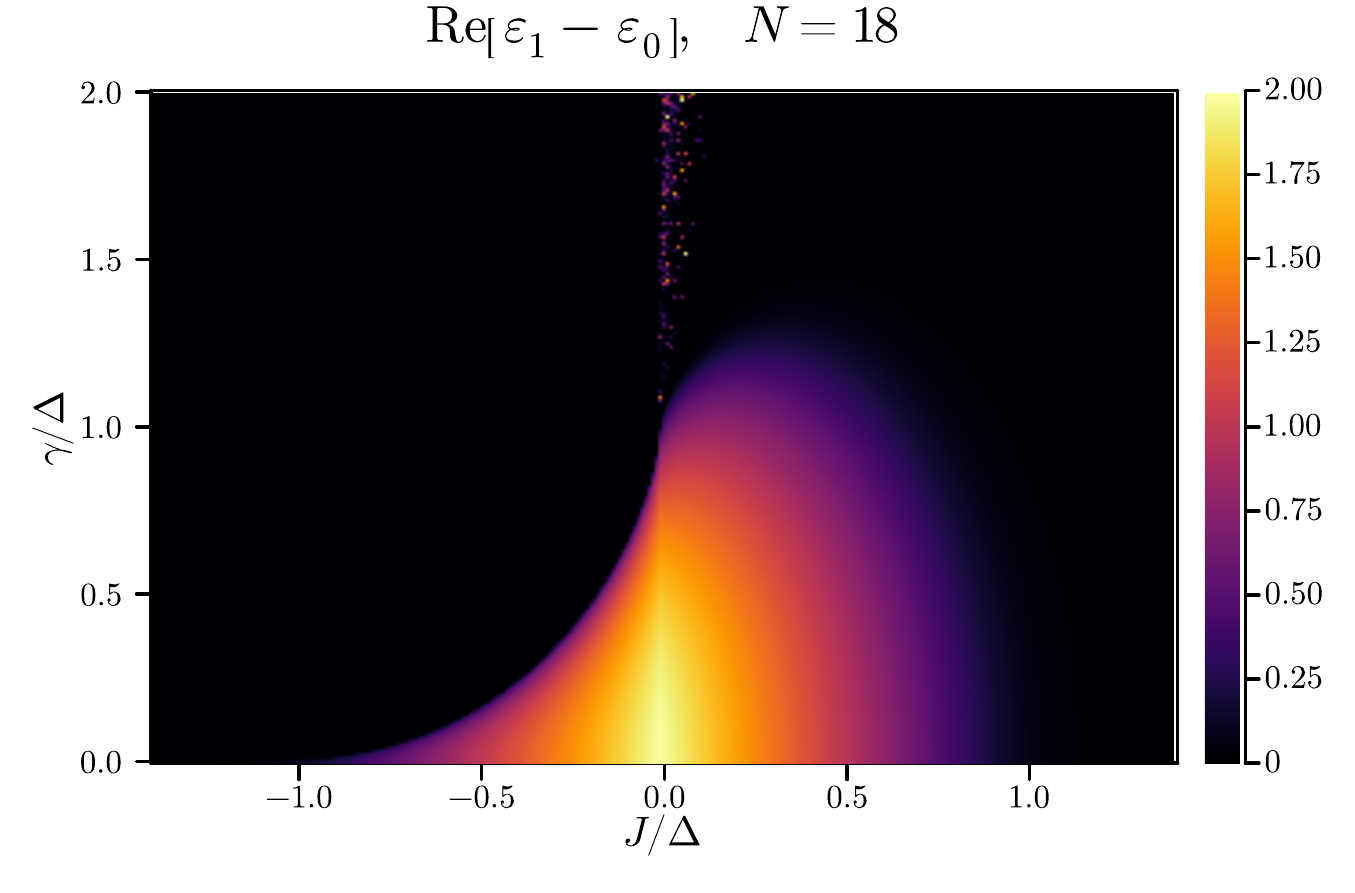}
\caption{Color plot of the real part of the energy gap, $\Re e (\Delta \varepsilon)$ between the first excited and the ground states for spin chains of $N=18$.}
\label{gap-colorplot}
\end{figure}

In the case of non-Hermitian $\BP\BT$-symmetry transverse field Ising  spin chains we numerically calculate the dependence of the real part of the gap, $\Re e (\Delta \varepsilon)$, on the parameters of $J/\Delta$ and $\gamma/\Delta$. For the spin chain of $N=18$ this dependence is presented in Fig. \ref{gap-colorplot} in the form of two-dimensional color plot. In agreement with Fig. \ref{st4} one can also conclude that the energy gap 
$\Delta \varepsilon$ has an imaginary part for an antiferromagnetic coupling ($J<0$) only, indicating the presence of a sharp transition between the $\BP\BT$-symmetry broken macroscopic antiferromagnetic  and  $\BP\BT$-symmetry preserved paramagnetic states. Concluding this subsection we stress that for spin chains with a ferromagnetic coupling ($J>0$) the energy gap \textit{does not} contain the imaginary part for a whole range of parameter $\gamma$, indicating the presence of quantum phases preserving the  $\BP\BT$-symmetry. 
Notice here that the energy gap $\Re e ( \Delta \varepsilon)$ demonstrates scarce changes in a whole range of parameters $J$ and $\gamma$ as a total number of spins  varies from $N=10$ up to $N=18$.  
%\begin{figure}[t]
%\includegraphics[width=\textwidth]{gap-scaling.pdf}
%\caption{Scaling of gap.}
%\label{gap-colorplot}
%\end{figure}

\subsection{Spatial correlations of the local spin polarization in the ground state.\label{section:corfuncs}}
\begin{figure}[t]
\includegraphics[width=\textwidth]{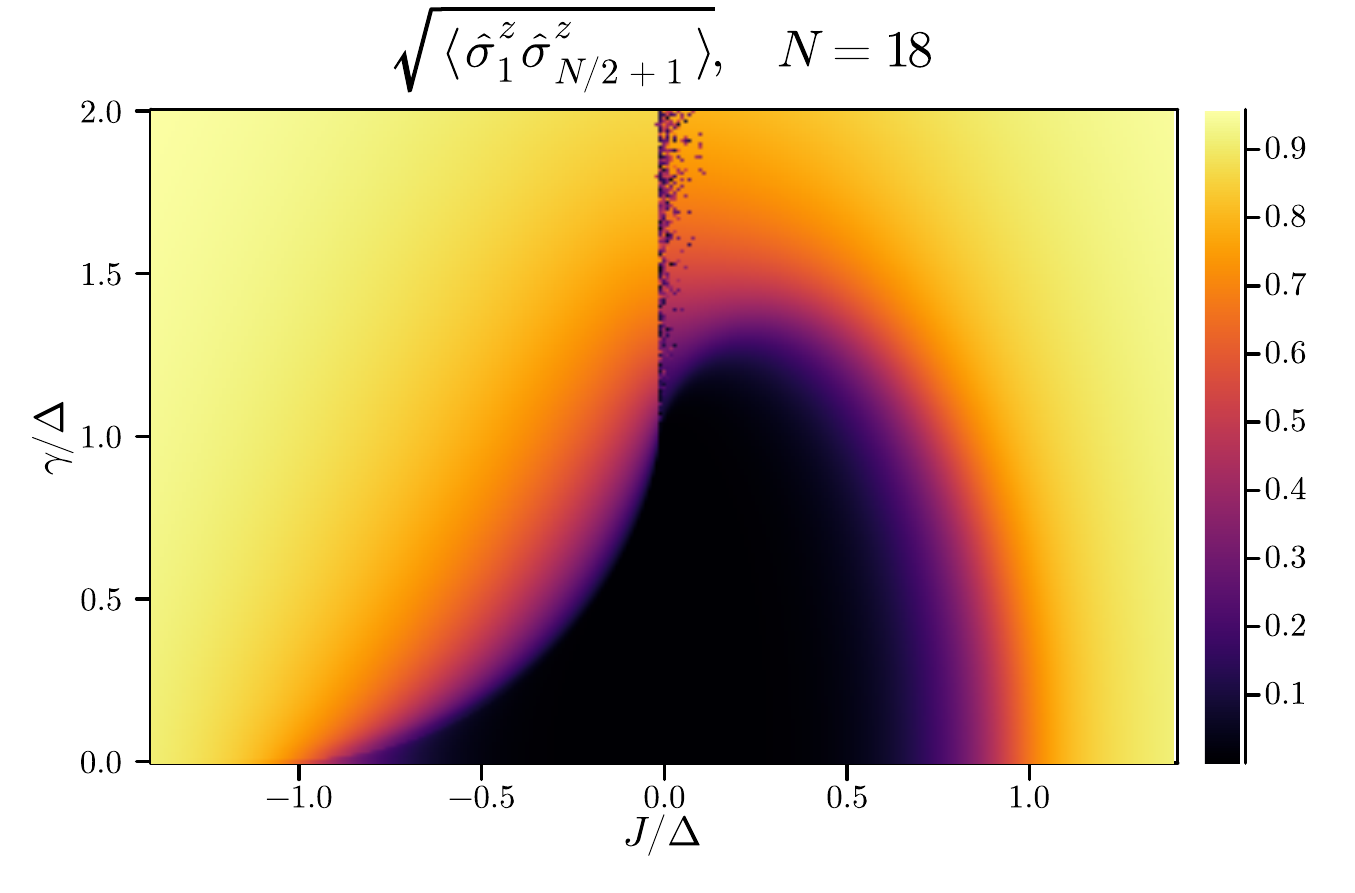}
\caption{Color plot of the spatial correlations characterized by the square root of the midpoint correlation function $\sqrt{C(|1-(N/2+1)|)}$ for spin chains with $N=18$. 
%parameter for $N=18$. 
%To estimate the order parameter, we computed  %$\sqrt{\langle\hat\sigma_1^z\hat\sigma_{N/2+1}^z\rangle}$.
}
\label{op-colorplot}
\end{figure}

Quantum phase transitions are identified more precisely by analyzing the spatial correlations of physical quantities. For chains of interacting spins it is convenient to use spatial correlation function of the local spin polarization ($z$-component of the magnetization), see Eq. (\ref{corrfunction}). In Fig. \ref{op-colorplot} we present a two-dimensional color plot of the dependence $\sqrt{C(|1-(N/2+1)|)}$ on the parameters $J/\Delta$ and $\gamma/\Delta$ for a spin chain with $N=18$. This characteristics $\sqrt{C(|1-(N/2+1)|)}$ can be considered as the order parameter which goes to zero value in the $\BP\BT$- symmetry preserved paramagnetic state. The sharp lines determining the corresponding quantum phases (compare with Fig. \ref{gap-colorplot} ) are very well seen in Fig. \ref{op-colorplot}.  However, we obtain also a substantial dependence of the order parameter on a total number of spins $N$ as $N$ varies from $10$ to $18$, and therefore, to precisely determine the quantum phase transitions lines in the limit of $N \rightarrow \infty$ we apply the ordinary finite size scaling procedure \cite{pang2019critical,asada2002anderson,Sandvik-2010}.

%\begin{figure}[t]
%\includegraphics[width=\textwidth]{order-parameter-scaling.pd%f}
%\caption{Scaling of order parameter.}
%\label{gap-colorplot}
%\end{figure}

%\begin{figure}[t]
%\includegraphics[width=3.4in]{figures/convergence.pdf}
%\caption{Convergence of the end-to-end correlation function %$\langle\hat\sigma_1^z\hat\sigma_N^z\rangle$ for the chains %of different size $N$. The curve at $N=\infty$ is the %analytic result of Lieb et. al. \cite{Lieb-61}.}
%\label{conv}
%\end{figure}

\subsection{The correlation length: finite size scaling.\label{section:fss}}

\begin{figure}[p]
\includegraphics[width=\textwidth]{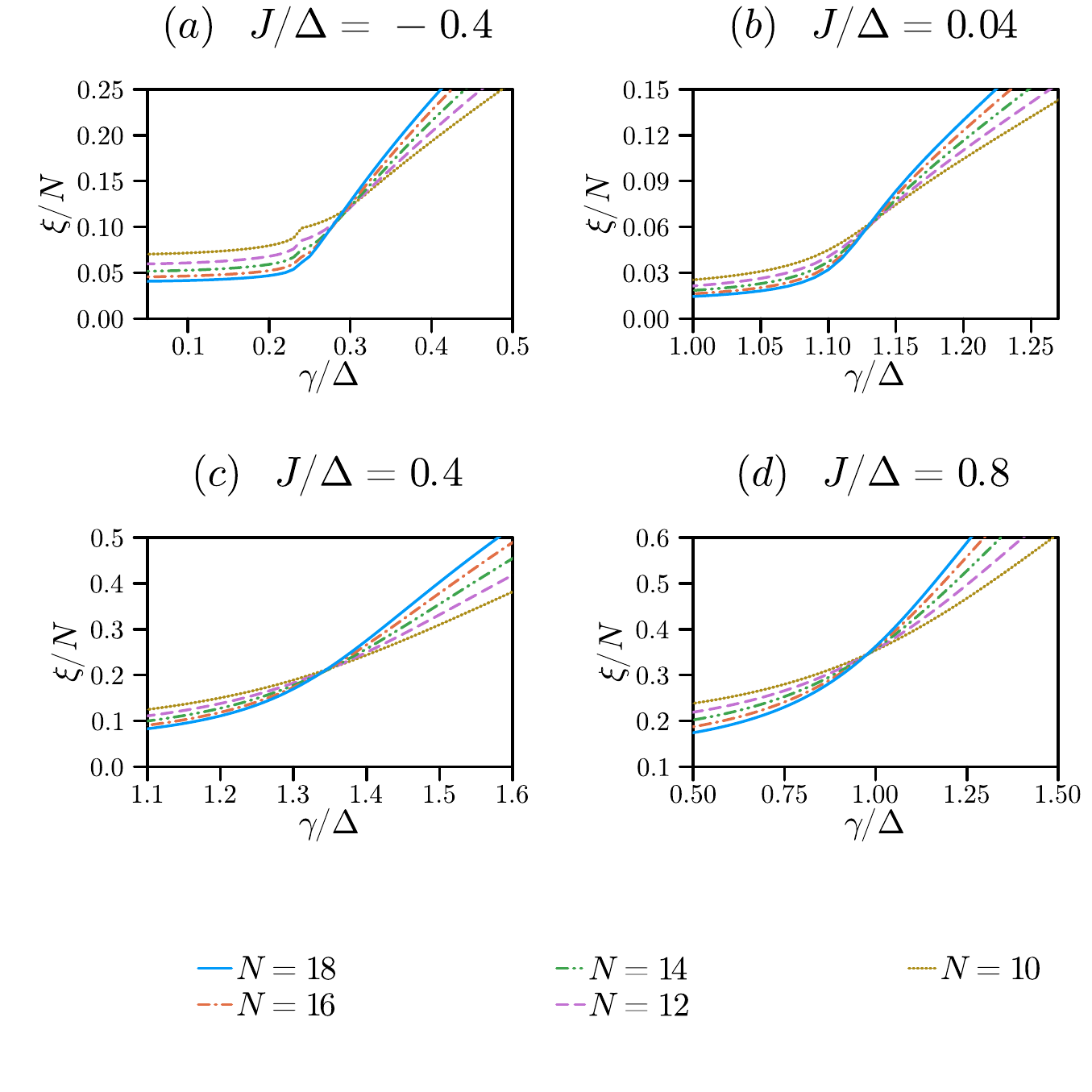}
\caption{The curves $\xi/N$ vs. $\gamma/\Delta$ for fixed values of $J$ demonstrating the scaling behavior, are shown. The values of $J$ were chosen as $-0.4$(a), $0.04$(b), $0.4$(c) and $0.8$(d). }
\label{scalingCorrLength-colorplot}
\end{figure}

As the practical way to compute the correlation length $\xi$, we adopt the definition based on the Fourier transform of the correlation function $C(j)$ (see Ref.~\cite{Sandvik-2010}):
\begin{equation}
    \xi = \frac{1}{q_1}\sqrt{\frac{S(0)}{S(q_1)} - 1},
    \label{correlation-length}
\end{equation}
where $q_1 = 2\pi/N$ and
\begin{equation}
    S(q) = \sum_{j=0}^N \cos{qj} C(j).
    \label{S-definition}
\end{equation}
In the case of the antiferromagnetic couplings ($J<0$), we compute the Fourier transform of the absolute value of the correlation function $C(j)$. The correlation length $\xi$ depends on the parameters $J$ and $\gamma$ as well as the total number of spins $N$. However, the dependence of the ratio $\xi/N$ on $\gamma$ for a fixed value of $J$ demonstrate a standard scaling behavior \cite{pang2019critical,asada2002anderson,Sandvik-2010}, i.e., all curves for different values of $N$ intersect in a single point. It is presented in Fig. \ref{scalingCorrLength-colorplot} for a few values of $J$.

\section{Phase diagram of non-Hermitian  $\BP\BT$-symmetric transverse-field Ising spin chain.
    \label{section:phasediag}}

\begin{figure}[p]
\includegraphics[width=\textwidth]{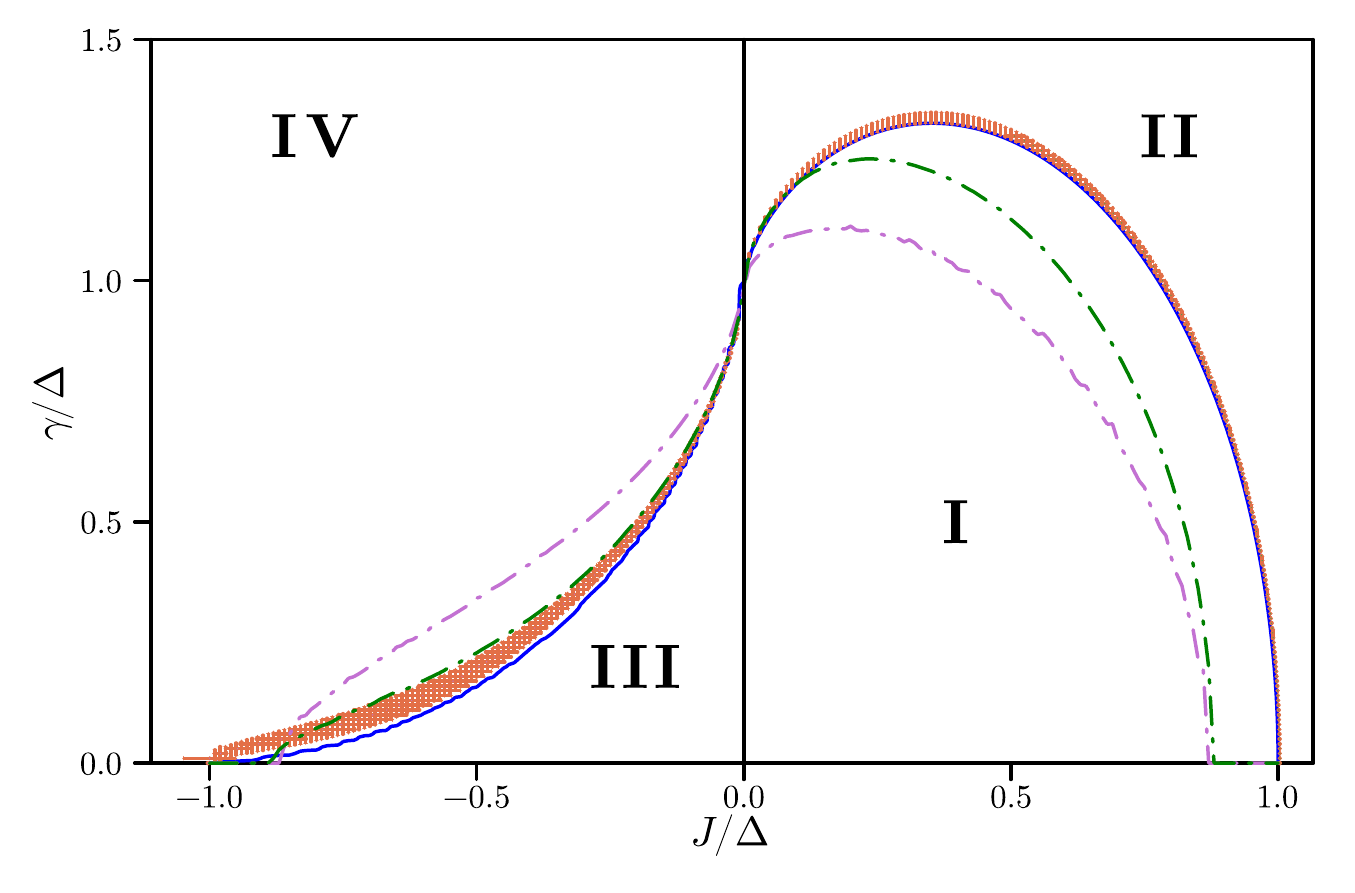}
\caption{Quantum phase diagram of the ground state. Orange points with error bars determine the critical line obtained using the intersection points of both $\xi/N$-$\gamma/\Delta$ and $\xi/N$-$J/\Delta$ curves with different $N$ (see Fig. \ref{scalingCorrLength-colorplot}). Blue solid line is the critical line determined by the threshold level curve for the gap.  Purple and green dash-dotted lines are the critical lines obtained using the Bethe-Peierls approximation with 2-spin and 6-spin clusters, respectively (see section~\ref{section:BP}). The obtained quantum phases are denoted as follows: 
%$\BP\PT$-symmetry preserved 
quantum paramagnetic ($\mathbf{I}$) and  ferromagnetic ($\mathbf{II}$) phases ; $\BP\BT$-symmetry preserved quantum paramagnetic ($\mathbf{III}$) and $\BP\BT$-symmetry broken ($\mathbf{IV}$)  antiferromagnetic phase.}
\label{pd}
\end{figure}

By making use of the dependencies of the energy gap $\Delta \varepsilon$ (see Fig. \ref{gap-colorplot}), the order parameter $\sqrt{C(|1-(N/2+1)|)}$ (see Fig. \ref{op-colorplot}) and the coherence length $\xi$ (see Fig. \ref{scalingCorrLength-colorplot}) on the parameters $J$ and $\gamma$ we obtain a \textit{complete quantum phase diagram} presented in Fig. \ref{pd}. 
%To conclude this section 
We stress here that the quantum phase transition line occurring for $J<0$ is the exception line separating the $\BP\BT$-symmetry preserved paramagnetic ($\mathbf{III}$) and $\BP\BT$-symmetry broken ($\mathbf{IV}$)  antiferromagnetic phases. In fact, close to the quantum phase transition line such non-Hermitian spin chain with an antiferromagnetic coupling can be qualitatively described as a single macroscopic two-level system, where two even (odd) sub-lattices form corresponding macroscopic eigenstates, in the presence of a global gain (loss) $\gamma$. 
In the presence of a ferromagnetic coupling ($J>0$) the $\BP\BT$-symmetry of the ground and first excited states is preserved, and in a narrow region of $\gamma > 1$ we obtain \textit{two} quantum phase transitions, i.e., ferromagnet-quantum paramagnet-ferromagnet transitions: one is in the region of small $J$ and other one is in the region of large $J$.  

Such complex quantum phase diagram can be qualitatively obtained by using the Bethe-Peierls approximation (see Fig. \ref{pd}, purple and green dashed-dotted lines).

\section{Phase diagram in the Bethe-Peierls approximation. \label{section:BP} }
%\section{Bethe-Peierls Approximation}
It is instructive how qualitatively correct results may be obtained by employing the Bethe-Peierls approximation~\cite{Bethe-1935, Peierls-1936, Weiss-1948}. 
%The idea of the method is quite simple. Let me illustrate it %in the absence of non-local repulsion (see also).
Generally in this approach one has to choose a central spin (marked as $0$) and treat its  interactions with the adjacent spins exactly, and  at the same time the two neighbouring spins are lumped together into a single spin (marked as $1$). The interaction of the nearest neighbours with next-to-nearest neighbours is taken into account by introducing the internal effective magnetic field determined by the average $z$-component of the magnetization, $M$. Diagonalizing the effective $4 \times 4$ Hamiltonian $\hat H_\mathrm{eff}$ we obtain the ground state $|GS(M)\rangle$.
The effective Hamiltonian has to be accompanied by the self-consistency condition written as
\begin{equation}
 \langle GS(M) |\sigma_0^z |GS(M)\rangle = \langle GS(M) |\sigma_1^z|GS(M)\rangle.\label{sce-bp0}
\end{equation}

Next we apply such generic procedure for  non-Hermitian $\BP\BT$-symmetry Ising spin chains. We will treat spins chains with antiferromagnetic and ferromagnetic couplings separately. As $J<0$ applying the $\pi$ rotation of all \textit{even} spins in $z-y$ plane we transform an antiferromagnetic spin chain with a staggered imaginary magnetic field into a ferromagnetic spin chain with a uniform imaginary field, and arrive to the effective Hamiltonian:
%at next-to-nearest neighbours, i.e., on the mean field level:
%of next-to-nearest neighbours. 
\begin{equation}
 \hat H_\mathrm{eff}^\mathrm{af} = \Delta\left(\sigma_0^x + 2\sigma_1^x\right) - 2|J|\sigma_0^z\sigma_1^z - 2|J|M\sigma_1^z +i\gamma\left(\sigma_0^z + 2\sigma_1^z\right). \label{af-BP}
\end{equation}
Diagonalizing such Hamiltonian we obtain the ground state eigenvalue $E_{gs}(M)$ and eigenfunction, $|GS_R(M) \rangle$. Using the self-consistency equation (\ref{sce-bp0}), where 
we need to replace $|GS(M)\rangle$ with $|GS_R(M)\rangle$, we obtain the effective magnetization $M$, and obtain the quantum phase transition between the $BP\BT$-symmetry preserved paramagnetic ($M=0$) and broken antiferromagnetic ($M \neq 0$) states. The quantum phase transition line is shown in the left part of Fig. \ref{pd} by purple dashed-dotted line.

For spin chains with a ferromagnetic coupling ($J>0$) the effective Hamiltonian for an \textit{odd} central spin $0$ acquires the form
\begin{equation}
 \hat H_\mathrm{eff}^\mathrm{f} = \Delta\left(\sigma_0^x + 2\sigma_1^x\right) - 2J\sigma_0^z\sigma_1^z - 2JM\sigma_1^z +i\gamma\left(\sigma_0^z - 2\sigma_1^z\right). \label{}
\end{equation}
Notice here that the effective Hamiltonian for an \textit{even} central spin is obtained by complex conjugation, and therefore, the ground state for even spin is obtained by complex conjugation from the ground state for odd spin. Moreover, the averages $\langle GS_R(M) |\sigma_0^z |GS_R(M)\rangle$ and $\langle GS_R(M) |\sigma_0^z |GS_R(M)\rangle$ are unaltered if the state is complex conjugated. 
%Physically, it means that we expect the same average magnetization at odd and even sites. 
As a result, we can still use Eq.~\eqref{sce-bp0} as the self-consistency condition.

Repeating the procedure presented above for antiferromagnetic spin chains, we obtain the effective magnetization $M$ for $J>0$, and obtain the quantum phase transitions between the $BP\BT$-symmetry preserved paramagnetic ($M=0$) and ferromagnetic ($M \neq 0$) states. The quantum phase transition line is shown in Fig. \ref{pd} by purple dash-dotted line. To conclude this section we notice that for a fixed value of $\gamma >1$ the Bethe-Peierls approximation provides two quantum phase transitions, i.e. ferromagnet-paramagnet-ferromagnet, that is in a good accord with numerically exact calculations.
We also obtain that substituting two central spins $0$ and $1$ on the spin cluster composed of $6$ interacting spins allows to substantially improve the agreement. In this case, the self-consistency equation is obtained by comparing the magnetization of the two central spins with the magnetization of the two next-to-central ones. The resulting phase transition line is shown in Fig.~\ref{pd} by green dash-dotted line.

\section{Conclusions. \label{section:conclusions}}

In conclusion we theoretically study various quantum phases occurring in non-Hermitian $\BP\BT$-symmetric  Ising spin chains in the transverse magnetic field, $\Delta$. Non-Hermitian part of the Hamiltonian is provided by imaginary staggered longitudinal magnetic field. Physically this model describes also one-dimensional chains of interacting superconducting qubits in the presence of staggered gain (loss) $\gamma$. The presence of a particular  quantum phase is determined by an interplay of two parameters, the interaction strength between adjacent spins $J$ and the gain (loss) $\gamma$. 

Using the direct numerical diagonalization of the Hamiltonian (\ref{Ising-Ham}) for Ising chains composed of $N$ ($N$ was up to $18$) interacting spins, accompanying by the scaling procedure for the correlation length $\xi(N)$, we were able to construct the complete quantum phase diagram $\gamma/\Delta$-$J/\Delta$ in the thermodynamic limit $N \rightarrow \infty$. The quantum phase diagram with the quantum phase transitions line is presented in Fig. \ref{pd}. We identify four different quantum phases, i.e., paramagnetic and ferromagnetic states, and $\BP\BT$-symmetry preserved paramagnetic and $\BP\BT$-symmetry broken antiferromagnetic states. These quantum phases differ by temporal (the energy gap between the first excited and ground states, $\Delta \varepsilon$, Fig. \ref{gap-colorplot}) and spatial (the order parameter, Fig. \ref{op-colorplot}) correlations.   

Overall one can see that the imaginary longitudinal magnetic field drives the quantum phase transition into the ordered phase. In particular, we obtain that spin chains with antiferromagnetic coupling ($J<0$) demonstrate the behavior resembling a macroscopic non-Hermitian $\BP\BT$-symmetry two-levels system (a single qubit). The quantum phase transition between the $\BP\BT$-symmetry preserved paramagnetic and broken antiferromagnetic states, occurs through the exception point. 
It is a result of the same parity between a staggered longitudinal magnetic field and  magnetizations of even/odd sublattices. At the same time, the model exhibits a 
%rather intriguing 
peculiar asymmetry between the antiferromagnetic and the ferromagnetic  types of interaction: the ground state on the ferromagnet-paramagnet transition does not go through the exception point. It is also interesting that spin chains with a ferromagnetic coupling ($J>0$) display a rather intriguing behavior, i.e., for fixed values of $\gamma >1$ an initial increase of the interaction strength $J$ as $J/\Delta \ll 1$ leads to the transition from the ferromagnetic state to the paramagnetic one. The paramagnetic state is stable in a wide region of $J/\Delta \simeq 1$. As $J > J_{cr}(\gamma)$ the system displays another quantum phase transition from the paramagnetic state to the ferromagnetic one. The detailed discussion of these two quantum phase transition, e.g., the critical indices of the dependencies of the order parameter $\sqrt{C(N/2)}$, coherence length $\xi$ and the energy gap $\delta \epsilon$ on the $J$ and $\gamma$ will be presented elsewhere. 

We also provide the qualitative description of the quantum phase diagram by making use of the Bethe-Peierls approximation, and obtain a good accord with numerically calculated quantum phase diagram. 

Finally, we notice that the obtained quantum phases and quantum phase transitions between them can be experimentally verified in $\BP\BT$-symmetric systems of interacting qubits by measuring e.g., the response to a small ac electromagnetic field \cite{tetling2022linear,blais2007quantum,jung2014multistability,fistul2022quantum}.

\textbf{Acknowledgements}
We thank Fl. S. Nogueira for fruitful discussions. We acknowledge the financial support through the European Union’s Horizon 2020 research and innovation program under grant agreement No 863313 'Supergalax'.\\

%Intuitively, this can be understood by the following argument: In the %absence of the dissipation, antiferromagnet chain can be mapped onto %the ferromagnet chain by performing a unitary transformation, which %flips the spins on all even sites. If one introduces then a staggered %longitudinal imaginary field, such a unitary transformation will turn %it into a uniform imaginary field. As such the imaginary component of %the energy is averaged to zero in the ground state of the ferromagnet, %while it adds up to a large value in the case of the antiferromagnet. %We corroborate this picture further by considering the mean-field description of the quantum phase diagram.

\bibliographystyle{elsarticle-num}
\bibliography{nonhermitian,ising}

\end{document}